\documentstyle[aaspp4]{article}

\newcommand{\mpc}{$h_{50}^{-1}$Mpc}

\begin{document}

\title{
A Study of Quasar Clustering at $z>2.7$ \\
from the Palomar Transit Grism Survey}

\author{
Andrew W. Stephens \altaffilmark{1} and Donald P. Schneider \altaffilmark{2}}
\affil{Department of Astronomy \& Astrophysics, Pennsylvania State University, University Park, Pennsylvania 16802}

\author{
Maarten Schmidt}
\affil{Palomar Observatory, California Institute of Technology, Pasadena, California 91125}

\author{
James E. Gunn \altaffilmark{3}}
\affil{Princeton University Observatory, Princeton, New Jersey 08544}

\author{David H. Weinberg}
\affil{Department of Astronomy, Ohio State University, Columbus, Ohio 43210}

\altaffiltext{1}{Current address:  Department of Astronomy, Ohio State University, Columbus, Ohio  43210}
\altaffiltext{2}{Guest Investigator, Palomar Observatory, California Institute of Technology.}
\altaffiltext{3}{Visiting Associate, Palomar Observatory, California Institute of Technology.}

\begin{abstract}

The quest for structure indicators at earlier and earlier times in the evolution of the universe has led to the search for objects with ever higher redshifts.  The Palomar Transit Grism Survey has produced a large sample of high redshift quasars ($z>2.7$), allowing statistical analysis of correlation between quasar positions.  In this study, clustering is identified through comparison with $100\,000$ Monte Carlo generated, randomly populated volumes, which are identical to the observed region in spatial coordinates, redshift distribution, and number of quasars.  Three pairs have been observed with comoving separations of 11.34, 12.97, and 24.13 \mpc\ (assuming $q_0=0.5$), smaller separations than would be expected to arise by chance in an unclustered distribution.  Selection effects are ruled out as a false source of clustering by scrambling the observed quasar coordinates and redshifts, which gives a pair separation distribution nearly identical to that of the Monte Carlo distribution.  Tests using the distribution of pair separations and nearest neighbor distances show that the observed pairs have a probability less than 0.1\% of arising in an unclustered distribution.  Using a maximum likelihood technique to estimate the correlation length $r_0$, assuming $\xi(r) = (r/r_0)^{-1.8}$, we find $r_0 = 35\pm15$ \mpc\ (comoving, $q_0=0.5$, 1$\sigma$ errors), a value much larger than the correlation length of present-day galaxies.

\end{abstract}

\section{Introduction}

The cosmic microwave background is remarkably uniform, but redshift surveys of the local galaxy distribution reveal coherent structures that extend over 100 \mpc\ or more ($h_{50} \equiv H_0/50\;{\rm km}\;{\rm s}^{-1}\;{\rm Mpc}^{-1}$).  One would like to trace the growth of this structure from the very early epoch probed by the microwave background through to the present day.  This objective has motivated clustering studies of homogeneous quasar samples, an approach pioneered by Osmer (1981).  In this paper, we analyze the clustering of 56 (of 90) high-redshift quasars ($2.7 < z < 4.75$) from the Palomar Transit Grism Survey (PTGS)(Schneider, Schmidt, and Gunn 1994, hereafter SSG), characterizing structure in a sample that contains some of the most distant quasars discovered to date.

When Osmer (1981) first obtained a survey of 174 quasars ($z < 3.5$), he was faced with the problem of how to quantify clustering, which he defined as density enhancements significantly in excess of expected random fluctuations.  After considering many statistical methods, he adapted and applied binning analysis, the nearest neighbor test, and the correlation function to his sample.  These tests dismissed several close groups picked out by eye as insignificant, and Osmer claimed no clustering on the 100--3000 \mpc\ scale, which was later confirmed by Webster (1982) using Fourier Power Spectrum Analysis.

Concerned that the lack of clustering evidence was an observational result of previous small sample sizes, Shaver (1984) devised a method of detecting clustering by comparing quasars close both in redshift and position to those with large differences in either redshift or position.  This method allows for the analysis of inhomogeneous quasar samples, since selection effects cancel out.  He applied his method to the $\sim 2000$ quasars in the V\'eron catalog (V\'eron-Cetty \& V\'eron 1984), revealing possible clustering similar to that of galaxies of the current epoch, which was later confirmed and quantified by Kruszewski (1988), who detected statistically significant clustering on scales up to 32 \mpc\ at low redshift but none at $z > 1.3$.

Later studies of quasar clustering attempted to achieve large sample sizes, while still maintaining homogeneity to minimize selection effects.  Shanks et al. (1987) used 170 quasars from UK Schmidt plates, complete to $z = 2.2$ and $B < 21$, and detected clustering on small scales ($r<20$ \mpc) but not on large scales ($20<r<2000$ \mpc).  Drinkwater (1988) found no clustering on scales of 20--200 \mpc\ in a sample of 862 quasars ($1.8 < z < 2.6$) from six UKST low dispersion objective prism plates.  Crampton, Cowley \& Hartwick (1989) discovered an unusually large group of 23 quasars at $z \sim 1.1$, but no statistical evidence for clustering in the remaining 91\% of the Canada-France-Hawaii Telescope grens survey.  Such large quasar groups (LQGs) have been studied by Komberg et al. (1996), who detected 12 LQGs with redshifts $0.5<z\le2$, and sizes from $\sim 140$ to $\sim 320$ \mpc.

Recently Shanks and Boyle (1994) have examined the clustering in three combined surveys with a total of 690 QSOs at $0.3<z<2.2$, claiming a $\sim 4 \sigma$ detection with a best value of $\bar{\xi}$(r$<20$ \mpc)=$0.93\pm0.34$ for the mean 2-point correlation function of pairs with comoving separation r$<20$ \mpc\ ($q_0$=0.5).  Kundic (1997) has also evaluated the clustering properties of 232 quasars from the PTGS and found significant clustering on scales less than 40 \mpc.

This study will concentrate on the high-redshift quasars ($2.7 < z < 4.75)$ from the PTGS.  The extreme distance of these quasars allows analysis of the universe when it was only about ten percent of its current age, providing the earliest observational link between the homogeneity of the universe at recombination, and the clumpiness of today's environs.  If quasar clustering is indicative of the large-scale distribution of their parent galaxy groups, quantifying clustering of these very ``young'' objects could help discriminate between various models of structure formation.

\section{Observations}

The PTGS was carried out with the Hale telescope, using the 4-Shooter camera/spectrograph (Gunn et al. 1987) in time delay-and-integrate (TDI) mode in order to obtain large numbers of low-resolution slitless spectra.  The total effective area of the survey is 61.47 square degrees, which is covered in six narrow strips $8.5'$ wide.  A computer algorithm was used to search the $\sim$600,000 grism spectra for the presence of emission lines.  The primary target was the Lyman $\alpha$ line, which appears in the survey bandpass at redshifts between 2.7 and 4.8.  A total of 928 emission-line objects were found.  Approximately 70\% of the objects were low-redshift emission-line galaxies; 90 of the sources were quasars at redshifts between~2.757 and~4.733.

This study of high redshift quasar clustering concentrates on two strips of the survey.  The ``MN'' strip spans right ascensions from $09^{\rm h} 12^{\rm m}$ to $16^{\rm h} 45^{\rm m}$ (with a $4^{\rm m}$ break caused by clouds), at a central declination of $+46^{\circ} 27'$, covering 9.98 square degrees and containing 24 quasars with $z>2.7$.  The ``QR'' strip spans right ascensions from $08^{\rm h} 09^{\rm m}$ to $17^{\rm h} 14^{\rm m}$, at a central declination of $+47^{\circ} 39'$, covering 11.91 square degrees, with 32 $z>2.7$ quasars.  These two strips were selected for analysis because they have the faintest effective flux limits for quasar detection (this limit is determined by a combination of seeing, exposure time, sky brightness, and throughput of instrument; see SSG).  While they cover but a third of the survey area, they contain nearly two thirds of the high redshift survey quasars (56 out of 90).  The remaining strips are too sparse to yield statistically significant results (positive or negative) in our tests, though we include them in our calculation of the correlation length in section 4.6 below.

\section{Numerical Methods}

To determine absolute distances between quasar pairs, the quasar's equatorial coordinates are converted to rectangular comoving coordinates using the quasar's redshift as a measure of its distance

\begin{equation}
r = \frac{c}{H_0} \times  2 \left ( 1 - \frac{1}{\sqrt{1+z}} \right ),
\end{equation}

\noindent assuming that $q_0 = \frac{1}{2}, (\Omega = 1)$. 

To investigate quasar clustering in this sample, the observed sample is compared with a large number of randomly populated Monte Carlo generated samples which match the actual sample in sky coordinates and redshift distribution.  The right ascension coordinate is chosen randomly between the upper and lower strip limits, except in the MN simulation.  While obtaining the direct imaging data for the MN strip, the passage of a moderately thick cloud through the field rendered the data between right ascensions $11^{\rm h} 34^{\rm m}$ and $11^{\rm h} 38^{\rm m}$ unusable (see SSG).  If the simulation created a quasar with a position in this region, the quasar was deleted and a new one selected to take its place.  The declination coordinate is selected through a fractional areas technique which equates a random number to the fraction of the area below a declination, yielding a random declination given by

\begin{equation}
\delta = \sin^{-1}[R (\sin \delta_2 - \sin \delta_1) + \sin \delta_1],
\end{equation}

\noindent where $R$ is a randomly chosen number between zero and one, and $\delta_1$ and $\delta_2$ are the declination limits of the test strips.  Redshifts are chosen randomly from a smooth model of the observed redshift distribution (Figure 1).  

\begin{figure}
\plotone{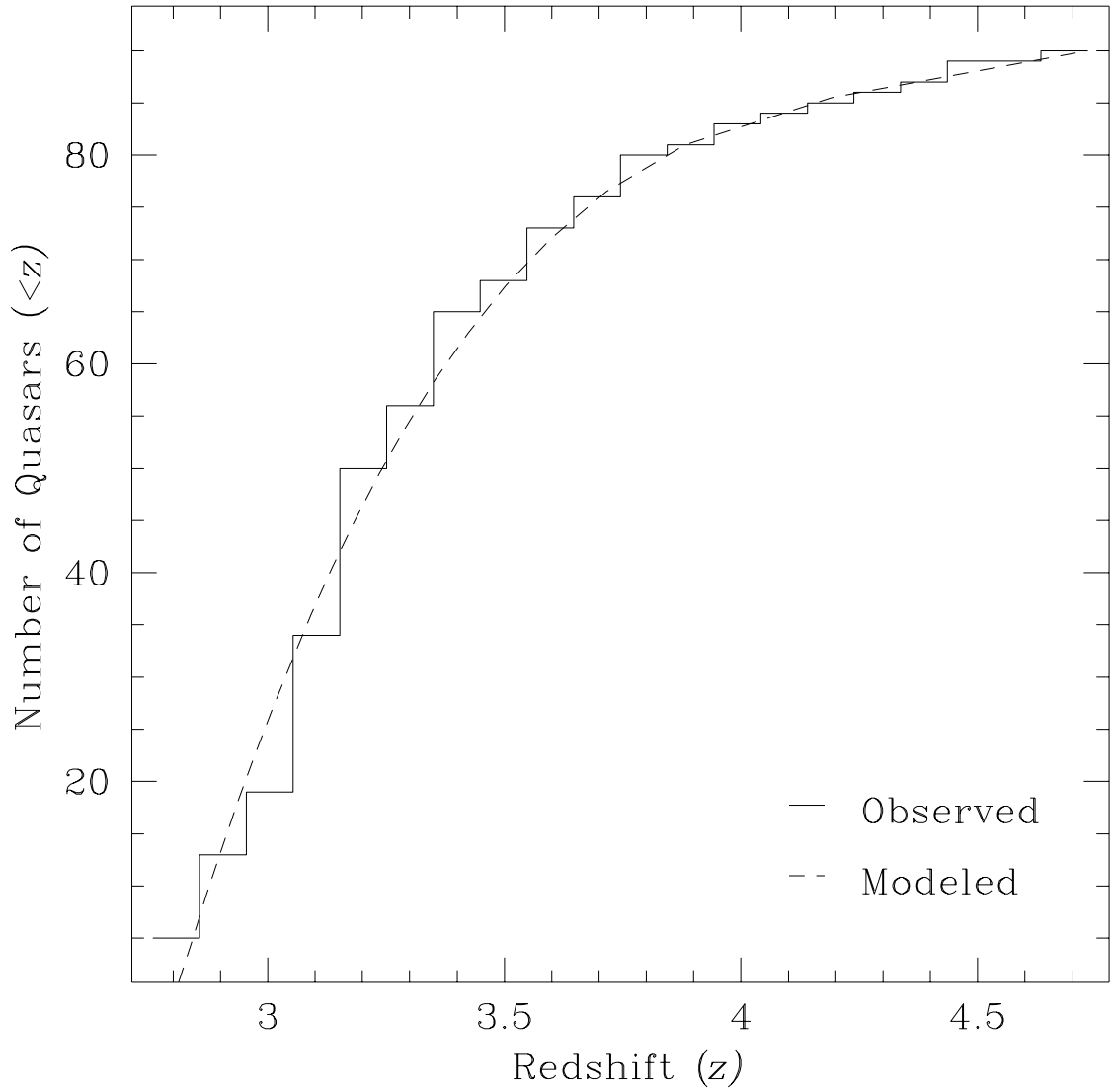}
\caption{Observed (solid) and modeled (dashed) PTGS redshift distribution.}
\end{figure}

All random numbers are generated by the subroutine RAN2 obtained from {\it Numerical Recipes} (Press et al. 1992), which was tested and found to be sufficiently random over the required number of calls.  These methods are used for each quasar in each trial, where the number of quasars randomly generated in each strip matches the observed number: 24 in the MN strip, and 32 in the QR strip.  Each strip is numerically modeled $100\,000$ times for the following statistical analyses.

\section{Tests for Clustering}

\subsection{Pair Separations} 

The first comparison is between the observed and average random distributions of separations between quasar pairs.  The distance is found between all possible quasars pairs in the volume, and these distances are placed into 25 logarithmic bins, where the bins are created to span nearly the entire range of observed separations, from 10 \mpc\ to approximately 8000 \mpc.

The average random pair separation distribution is found by adding each pair separation to the appropriate distance bin as each trial is executed, yielding the cumulative number of pairs in each bin from the entire $100\,000$ trials, which is then normalized by dividing by the number of trials.  The observed and average random quasar pair separation distributions are plotted for each of the two strips in Figures 2 and 3. 

These plots show that for separations greater than $\sim 60$ \mpc, the observed distribution of pair separations is very nearly random, while at separations $<30$ \mpc, quasar pairs exist where none should if the distribution were random.  The apparent deficit of observed pairs in the 20-100 \mpc\ bins of the QR strip is not statistically significant, as only 2.24 pairs are expected.  By contrast, the single pair in the smallest bin for this strip is quite significant; only 2.7\% of Monte Carlo trials have one or more pairs in this bin.  The two pairs with $r<30$ \mpc\ in the MN strip are much more significant still; only 0.523\% of the $100\,000$ Monte Carlo trials have two or more pairs in the first four bins.

\begin{figure}
\plotone{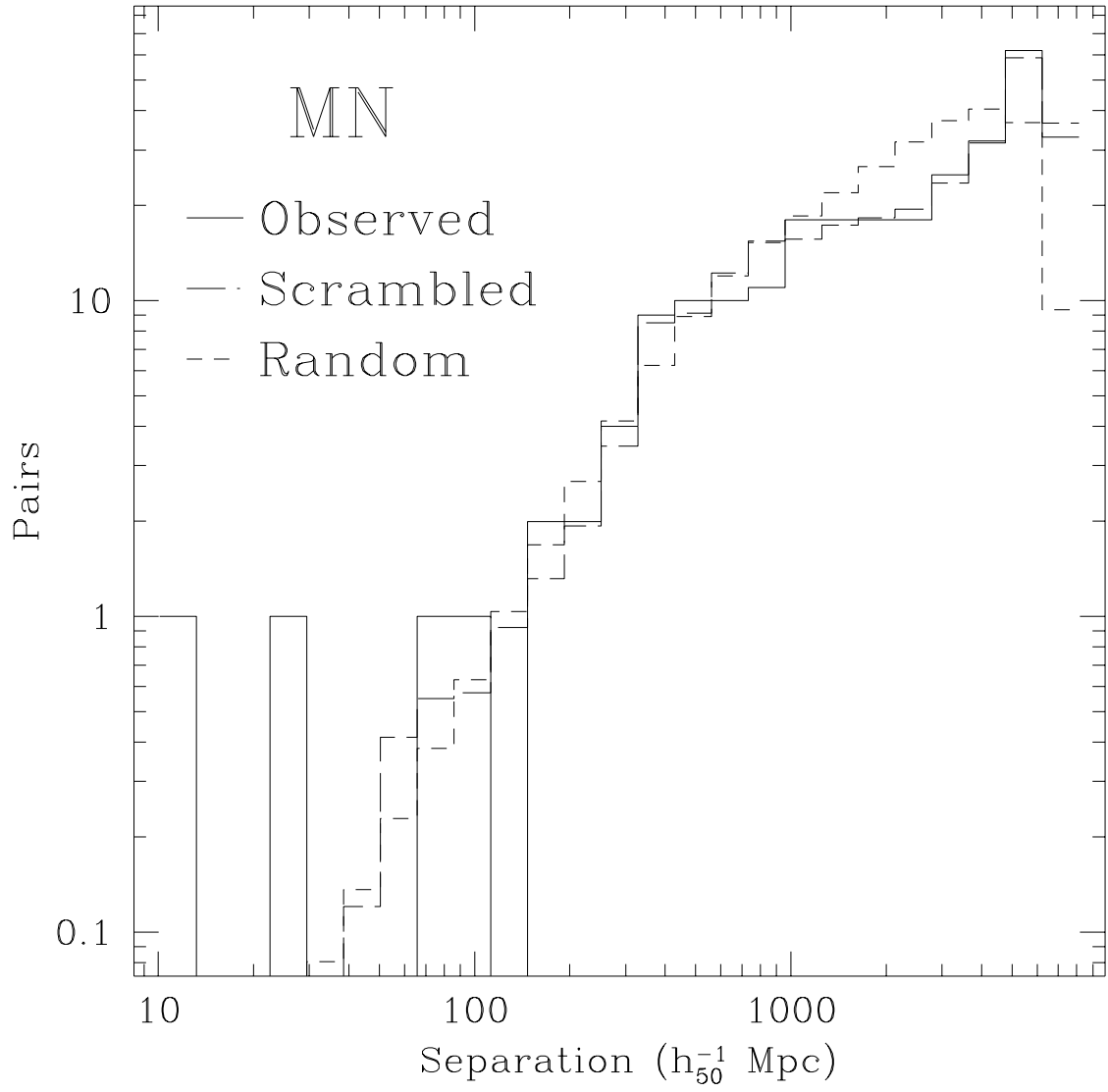}
\caption{MN Strip:  Observed (solid), Scrambled (long dashed) \& Random (short dashed) distributions.}
\end{figure}

\begin{figure}
\plotone{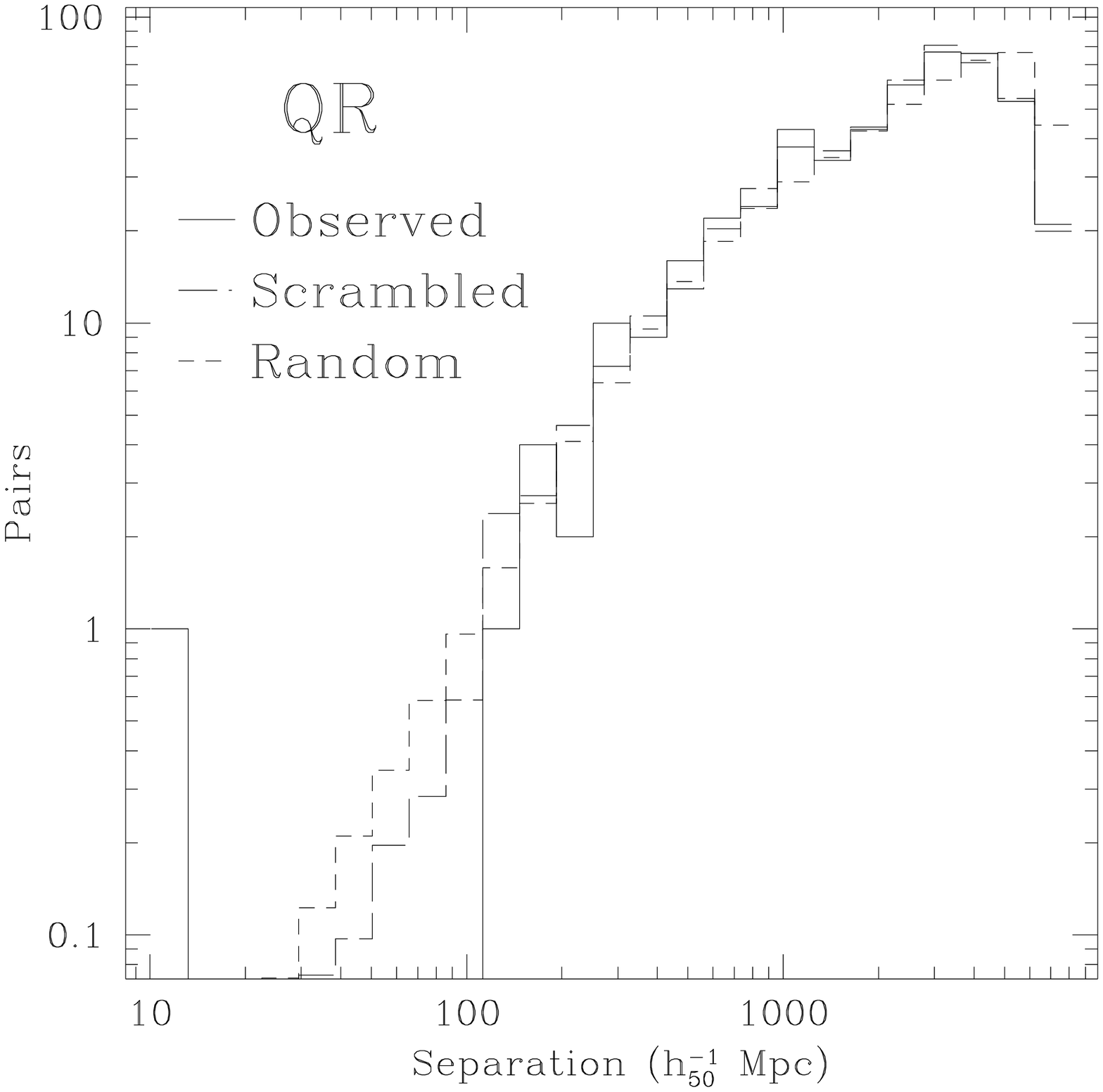}
\caption{QR Strip:  Observed (solid), Scrambled (long dashed) \& Random (short dashed) distributions.}
\end{figure}

\subsection{Scrambling} 

In order to determine if these excess quasar pairs are due to selection effects, the coordinates and redshifts of each of the observed quasars in the survey are scrambled and randomly redistributed $100\,000$ times, so that all coordinates are assigned to all redshifts in random combinations.  The resulting average scrambled pair separation distribution for each strip is also shown in Figures 2 \& 3.  

For the MN strip, the scrambled pair separation distribution very nearly matches the observed distribution on scales larger than $\sim 60$ \mpc, but scrambling does not produce significant numbers of close quasar pairs like the ones observed.  Similarly in the QR strip, the scrambled distribution is close to the observed distribution, except for in the smallest bin, where again, no significant number of close quasar pairs are evidenced like the one observed.  These results are nearly identical to the Monte Carlo generated results at the clustering scale under investigation, and show that the observed quasar pairs are {\em not} due to selection effects.

\subsection{Minimum Separations} 

For another test, one can ask how the minimum quasar pair separation in each of the randomly generated strips compares with the observed minimum quasar pair separation.  This is answered by Figure 4, which shows the cumulative probability distribution for any minimum separation in each quasar strip.  This distribution is obtained by determining the distance between the closest quasar pair in each of the $100\,000$ trials, placing them cumulatively into the standard logarithmic bins, and normalizing to one by dividing by $100\,000$.  Superimposed onto their respective curves are filled circles representing the closest observed quasar pairs for each strip, and the third quasar pair (with a separation of 24.13 \mpc) for reference.  This plot shows that the probability of obtaining {\it one} minimum separation as small as that exhibited in any of these three pairs is less than 10\%.

\begin{figure}
\plotone{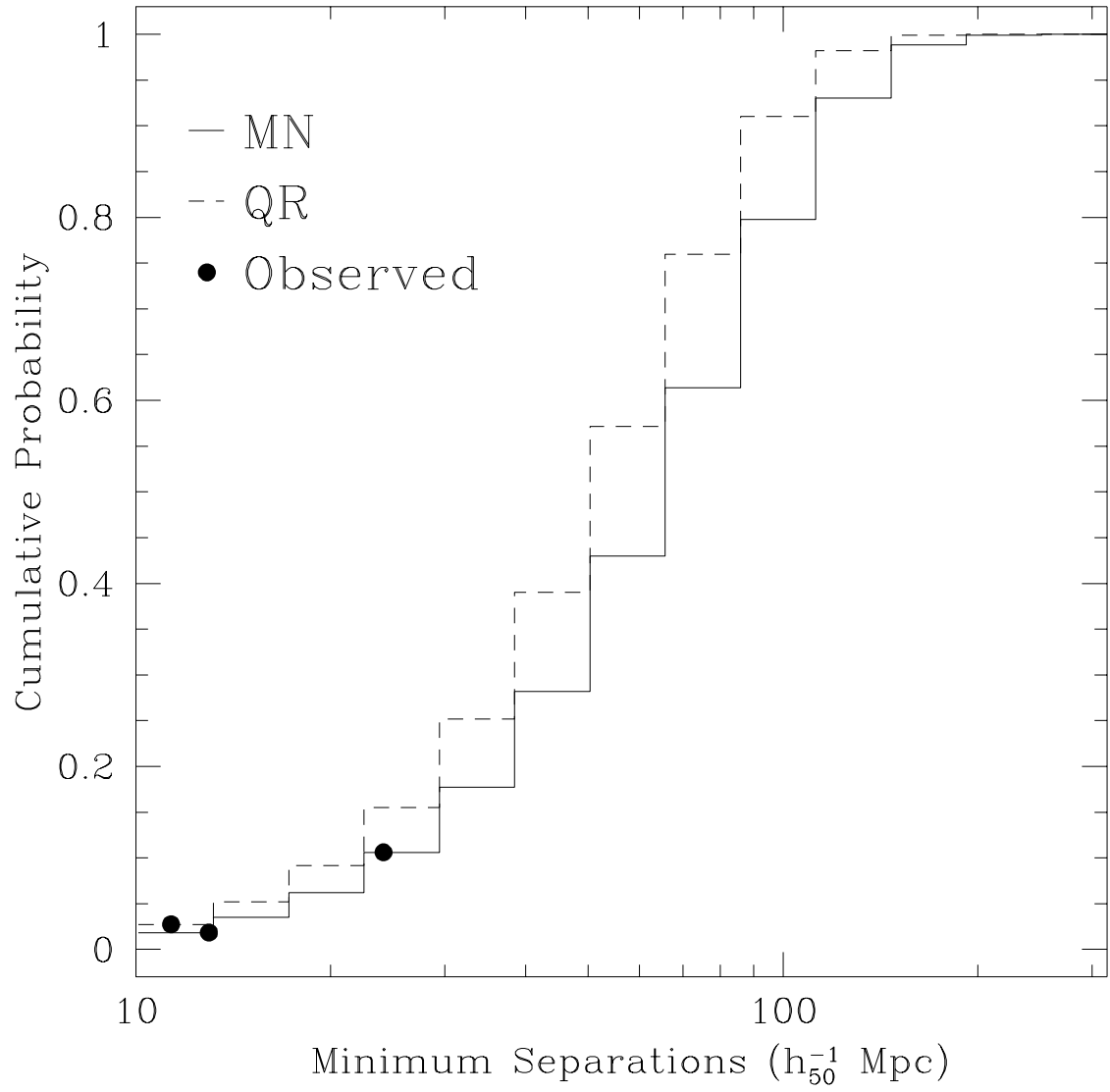}
\caption{Minimum random quasar pair separations; filled circles indicate observed quasar pairs.}
\end{figure}

\subsection{Nearest Neighbor Test} 

A nearest neighbor distance (the distance to the closest quasar) is calculated for each quasar in the survey, and these distances are placed into the same logarithmic bins.  For the randomly generated quasars, the nearest neighbor distance is calculated for each quasar in each trial, placed into bins, and the bins are averaged over the $100\,000$ trials,  yielding the mean number of quasar pairs in each bin.  This is compared to the observed number of nearest neighbor pairs in each bin in Figures 5 \& 6 for the MN and QR strips.  From these plots it is evident that for the MN strip, the nearest neighbor separation distribution peaks at $\sim 300$ \mpc, both for the randomly generated quasars and the observed quasars, {\em but} the observed nearest neighbor distribution is dramatically higher at very small separations ($<30$ \mpc), where the two quasar pairs lie (the plot registers two in each bin because a nearest neighbor distance was measured for {\em each} quasar, hence a single close pair will produce two identical nearest neighbor distances).  Both the observed and the random distributions for the QR strip also peaks near $\sim 300$ \mpc, but here again the observed distribution is elevated at the smallest distance bin, corresponding to the single pair in the QR strip.

\begin{figure}
\plotone{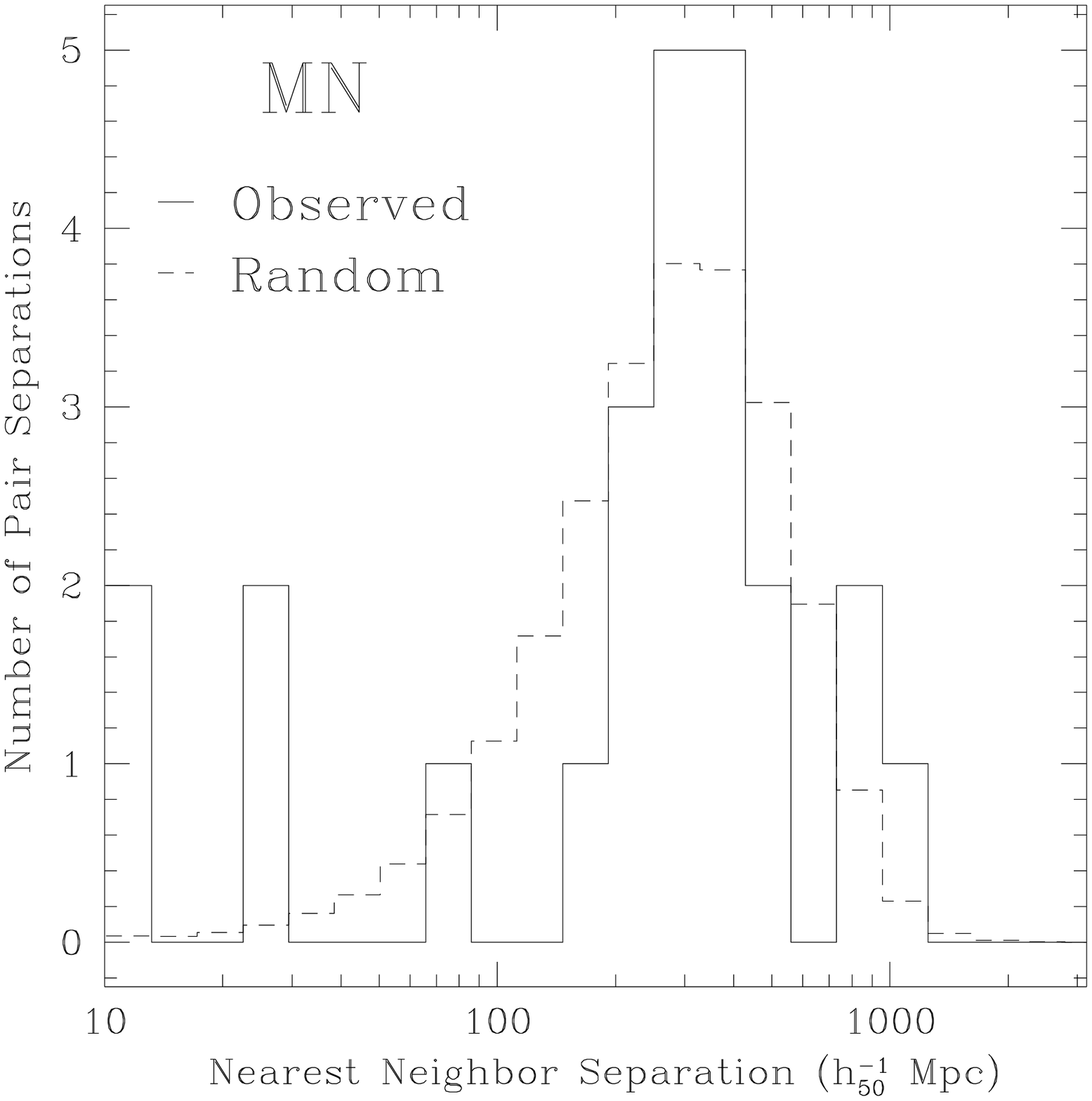}
\caption{Observed (solid) and mean random (dashed) nearest neighbor distribution of the MN strip.}
\end{figure}

\begin{figure}
\plotone{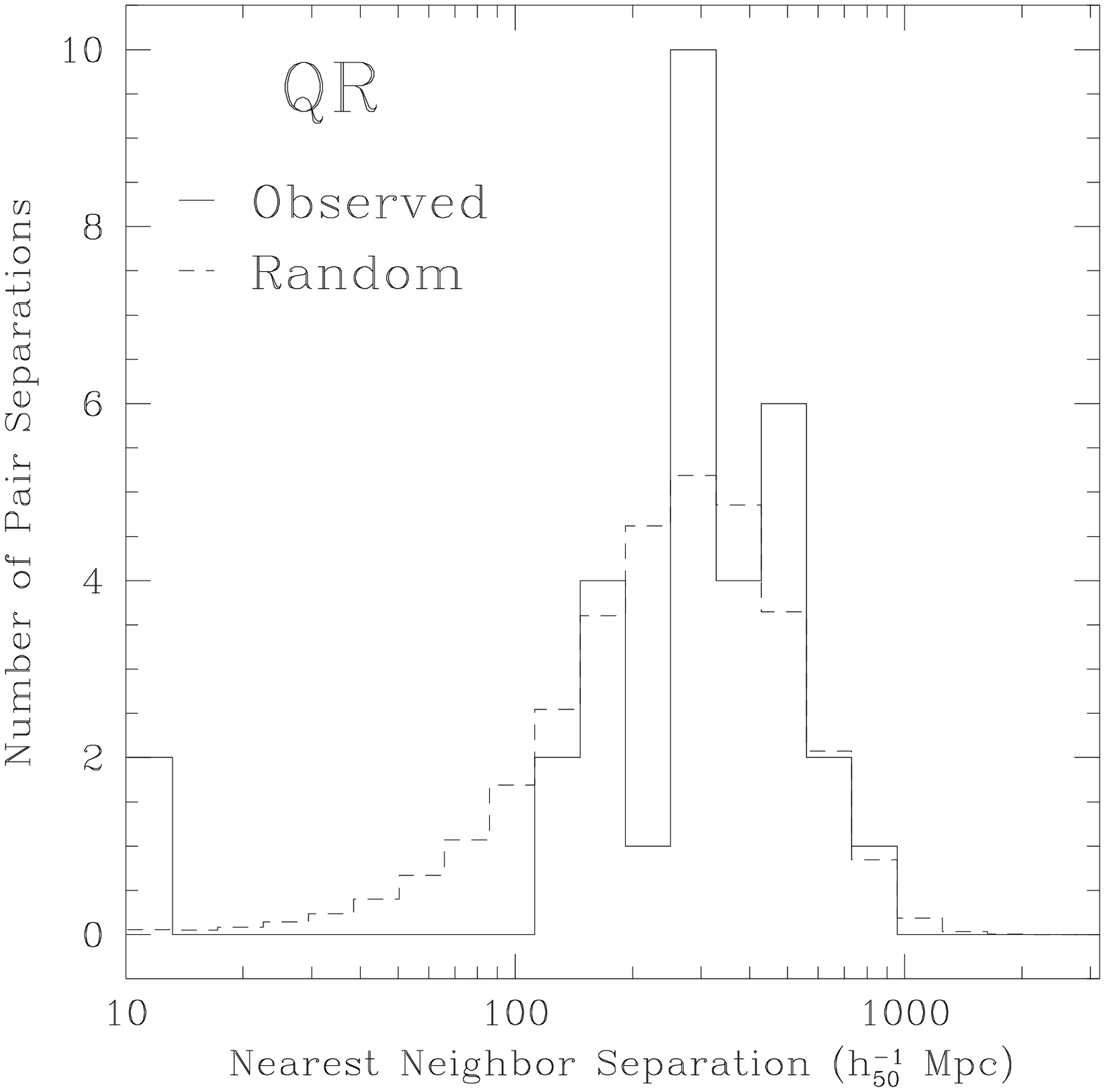}
\caption{Observed (solid) and mean random (dashed) nearest neighbor distribution of the QR strip.}
\end{figure}

\subsection{Probabilities of Observed Pair Separations}

\begin{deluxetable}{c r r r r r r r r} 
\tablecaption{Probabilities of obtaining pairs -- MN Strip}
\tablewidth{0pt}
\tablehead{
\colhead{Pairs} &  
\colhead{}	&
\colhead{}	&
\multispan4{Pair Separations in \mpc} \\
\colhead{}	&
\colhead{11.54} & 
\colhead{15.08} & 
\colhead{19.71} & 
\colhead{25.76} & 
\colhead{33.66} & 
\colhead{44.00} & 
\colhead{57.50} & 
\colhead{75.16} }
\startdata
0 & 98.158    & \bf 98.310 & \bf 97.226 & 95.319    & \bf 92.030 & \bf 87.320 & \bf 79.563 & 68.193     \\
1 & \bf 1.825 &  1.675     &  2.735     & \bf 4.570 &  7.644     & 11.840     & 18.190     & \bf 26.106 \\
2 &  0.017    &  0.014     &  0.038     &  0.110    &  0.317     &  0.803     &  2.079     &  4.997     \\
3 &  0.000    &  0.000     &  0.000     &  0.002    &  0.009     &  0.036     &  0.158     &  0.638     \\
4 &  0.000    &  0.000     &  0.000     &  0.000    &  0.000     &  0.001     &  0.009     &  0.061     \\
\enddata
\end{deluxetable}

\begin{deluxetable}{c r r r r r r r r} 
\tablecaption{Probabilities of obtaining pairs -- QR Strip}
\tablewidth{0pt}
\tablehead{
\colhead{Pairs} &  
\colhead{}	&
\colhead{}	&
\multispan4{Pair Separations in \mpc} \\
\colhead{}	&
\colhead{11.54} &
\colhead{15.08} &
\colhead{19.71} &
\colhead{25.76} &
\colhead{33.66} &
\colhead{44.00} &
\colhead{57.50} &
\colhead{75.16} }
\startdata
0 & 97.269    & \bf 97.452 & \bf 95.846 & \bf 92.986 & \bf 88.649 & \bf 81.413 & \bf 70.484 & \bf 56.109 \\
1 & \bf 2.693 &  2.515     &  4.067     &  6.762     & 10.681     & 16.742     & 24.654     & 32.424     \\
2 &  0.037    &  0.032     &  0.086     &  0.246     &  0.643     &  1.721     &  4.312     &  9.368     \\
3 &  0.000    &  0.000     &  0.001     &  0.006     &  0.026     &  0.118     &  0.503     &  1.805     \\
4 &  0.000    &  0.000     &  0.000     &  0.000     &  0.001     &  0.006     &  0.044     &  0.261     \\
\enddata
\end{deluxetable}

Assuming a Poisson distribution for the occurrence of a close quasar pair, it is possible to calculate the probability of such a pair.  The probability of obtaining $x$ pairs in any bin in the absence of clustering is

\begin{equation}
P(x) = \frac{\mu ^x e^{- \mu}}{x !},
\end{equation}

\noindent{where} $\mu$ is the mean number of pairs in that bin taken from the Monte Carlo simulations.  These probabilities are given as percentages for the first eight bins in Tables 1 and 2.  The probability corresponding to the observed number of quasar pairs in each bin is highlighted, revealing that the first and fourth bins of the MN strip, and the first bin in the QR strip are virtually the only bins which do \it not \rm conform to the predicted most probable number of pairs per bin.

\subsection{Correlation Function}

The most widely used descriptor of clustering is the correlation function

\begin{equation}
\xi(r) = \frac{N_{obs}(r)}{N_{ran}(r)} - 1,
\end{equation}

\noindent where $N_{obs}(r)$ is the observed number of pairs with separation $r$, and $N_{ran}(r)$ is the number of pairs expected for a random, unclustered distribution.  Thus, an unclustered distribution should yield $\xi(r) = 0$ on average.  With only three small-separation pairs, our estimate will necessarily be quite uncertain, but it is useful to present our results in this form to allow comparison with other estimates.

The histograms in Figures 7 \& 8 show $1+\xi(r)$ estimated in logarithmic bins using equation (4).  The long-dashed line shows the expectation $\xi(r)=0$ for an unclustered distribution.  The dotted lines indicate the one-sigma envelope for an unclustered distribution due to Poisson fluctuations in the number of pairs, where the standard deviation for each bin is determined from the distribution of quasar pairs from the Monte Carlo simulations,

\begin{equation}
\sigma = \sqrt{\frac{1}{n-1} \sum_{i=1}^n (x_i-\overline{x})^2 },
\end{equation}

\noindent where $\bar{x}$ is the mean number of pairs in the bin, and $x_i$ is the number in the $i^{th}$ Monte Carlo simulation.  The pair counts at large separations are consistent with Poisson fluctuations in an unclustered distribution, but the high estimates of $\xi(r)$ at $r<30$ \mpc\ are clearly inconsistent with a random quasar distribution.

\begin{figure}
\plotone{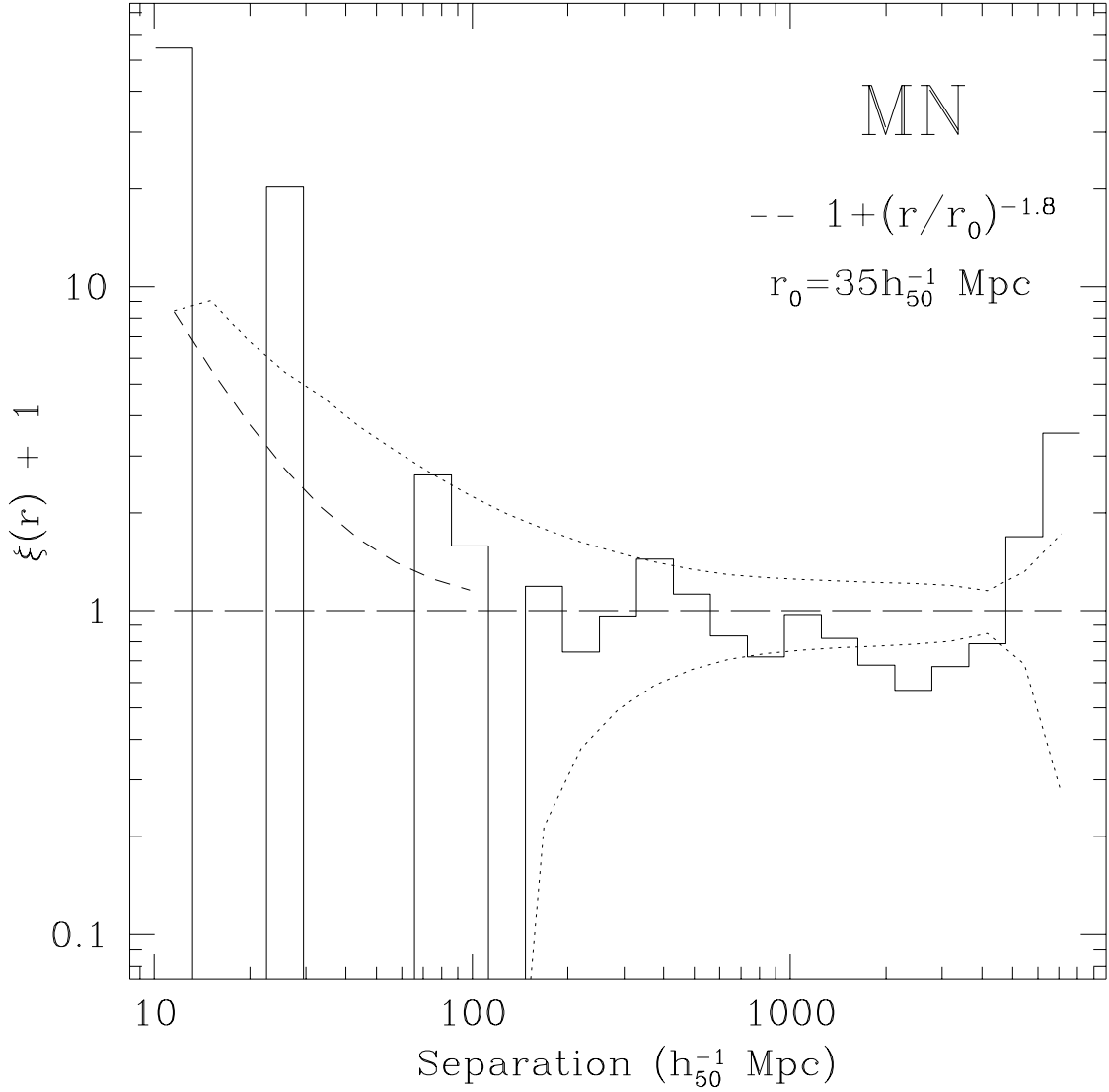}
\caption{Correlation function + 1 for the MN strip, the corresponding one sigma error envelope (dotted), and the predicted $\xi(r)+1$ assuming $r_0=35$ \mpc.}
\end{figure}

\begin{figure}
\plotone{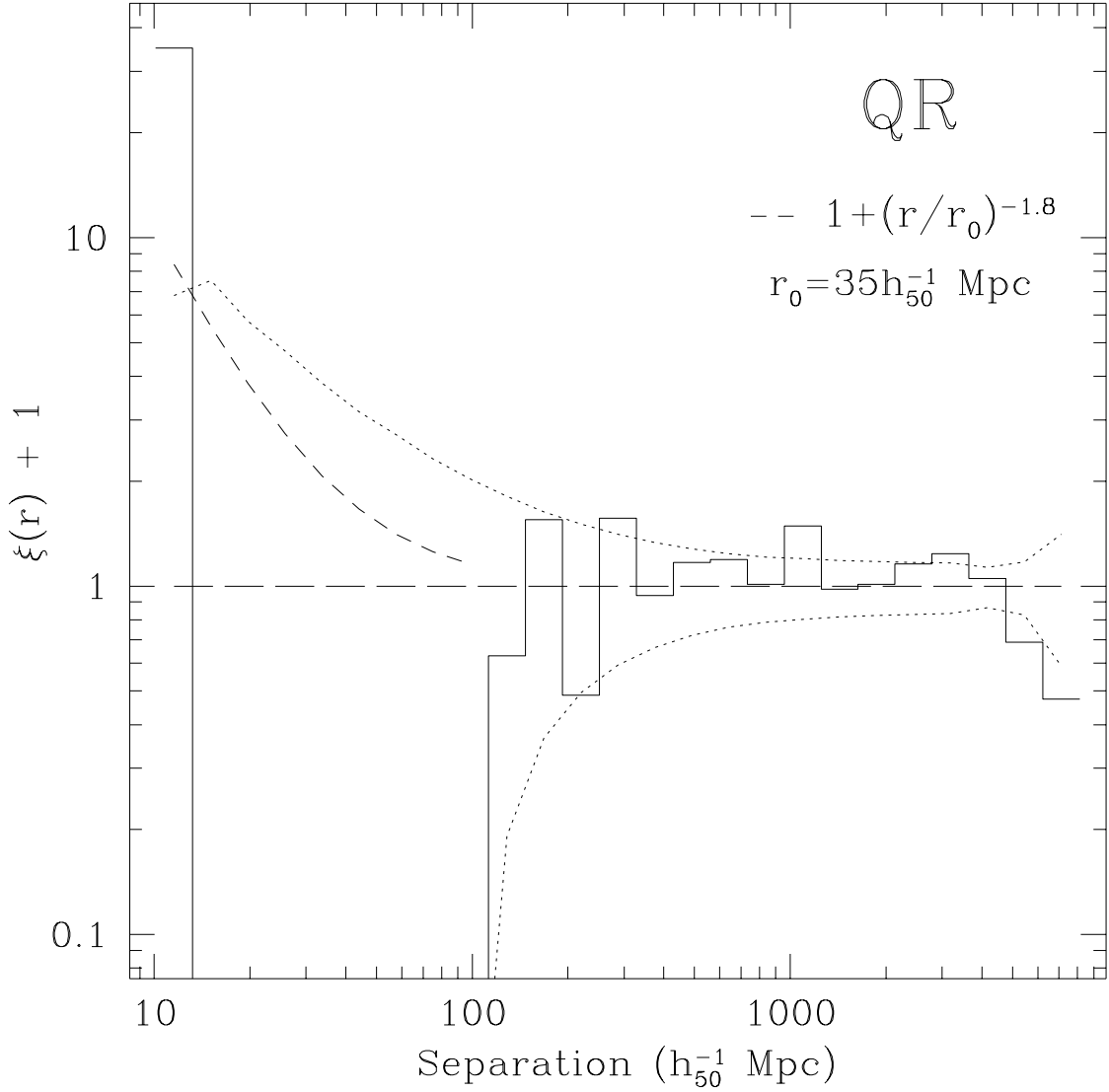}
\caption{Correlation function + 1 for the QR strip, the corresponding one sigma error envelope (dotted), and the predicted $\xi(r)+1$ assuming $r_0=35$ \mpc.}
\end{figure}

Because the quasar distribution is sparse, the probability of obtaining $N_{obs}$ pairs when $N_{ran}$ are expected is given by Poisson statistics (equation [3]).  This makes it simple to devise a maximum likelihood procedure for estimating parameters of the correlation function (Croft et al. 1997; Popowski et al. 1997).  Because our sample is so small we will assume that the true quasar correlation function has the same {\it form} as the present-day galaxy correlation function

\begin{equation}
\xi(r) = (r/r_0)^{-1.8},
\end{equation}

\noindent and estimate only the correlation length $r_0$ itself.

The calculation is performed simultaneously on all six strips of the PTGS.  Using 1 $h_{50}^{-1}$ Mpc bins, we sum the observed and random values of each strip for each bin, then find the value of $r_0$ that maximizes the likelihood of obtaining the observed number of pairs.  Figure 9 shows the results of this analysis, giving the maximum likelihood estimate $\hat{r_0}$ using the first $x$ bins (each 1 \mpc).  As the number of bins included in the calculation is increased, the number of quasar pairs increases sporadically.  With each new pair, the estimate $\hat{r_0}$ jumps, then declines smoothly until the next observed pair is encountered.

The $1 \sigma$ error in $r_0$ can be calculated from the second derivative of the likelihood function at its maximum,

\begin{equation}
\sigma = \left ( \frac{\partial^2 \ln L}{\partial r_0^2} \right )^{-1/2}.
\end{equation}

\noindent A more conservative error range, corresponding to $\sim 2 \sigma$ for Gaussian statistics, is given by finding the values of $r_0$ for which the likelihood falls to 10\% of its peak value,

\begin{equation}
\frac{ L(r_0) }{ L(\hat{r_0}) } = 0.1.
\end{equation}

\noindent The dashed and dotted lines in Figure 9 show these error ranges.  The estimate of $r_0$ depends on the number of bins we include in the calculation, but it is fairly stable for large bin numbers at $r_0 = 35 \pm 15$ \mpc\ (1 $\sigma$ errors).

\begin{figure}
\plotone{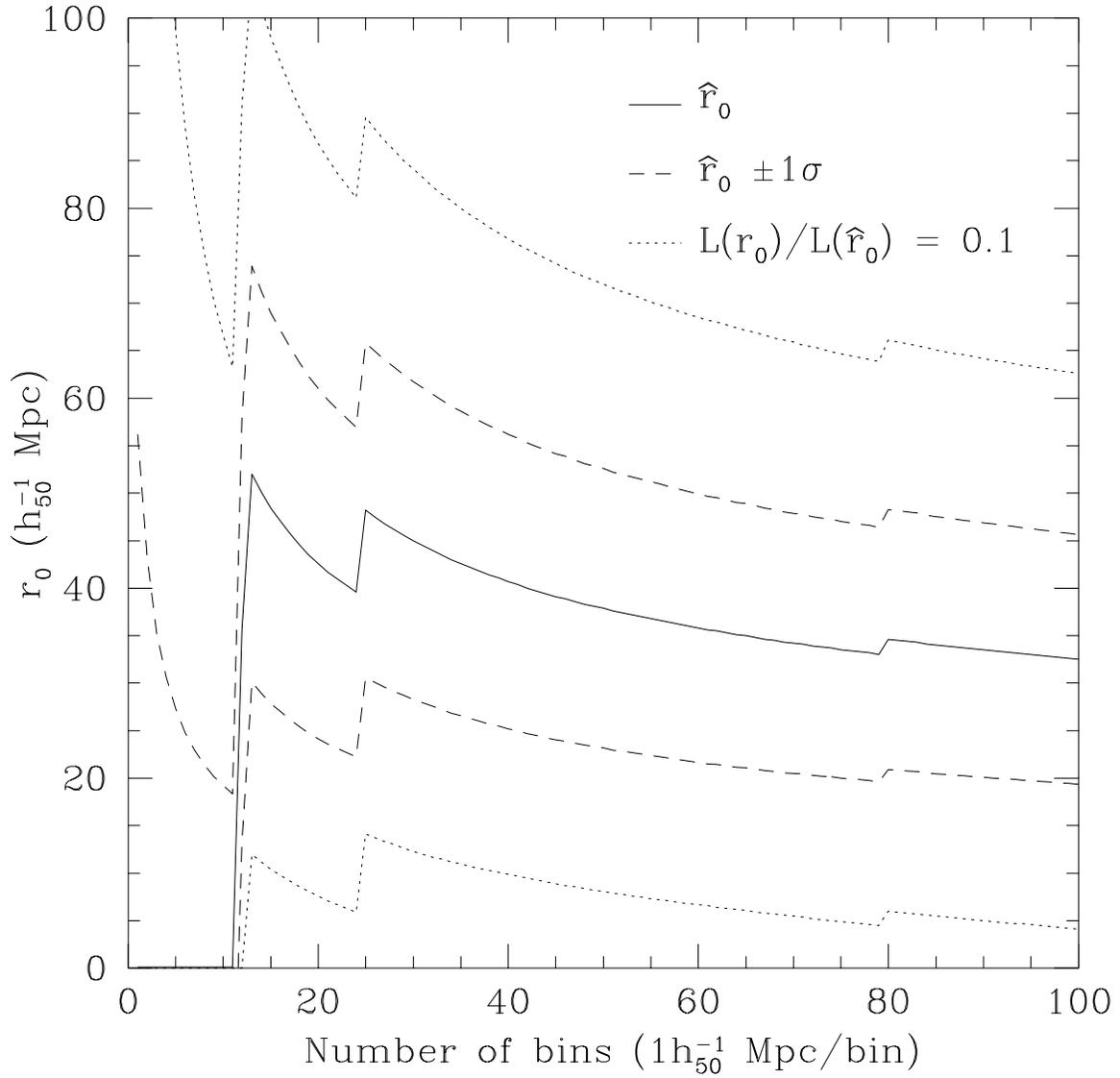}
\caption{Maximum likelihood estimate of the correlation length $r_0$ as a function of the number of 1 \mpc\ bins included in the calculation (solid line).  Dashed lines show the $1 \sigma$ range on $r_0$. Dotted lines show the range within which the likelihood exceeds 10\% of its peak value.}
\end{figure}

Our maximum likelihood estimate of $r_0$ is much larger than the value of $r_0 \sim 13$ \mpc\ estimated by Shanks \& Boyle (1994), though it is just consistent with this value at the $2\sigma$ level.  Given the small number of quasars in the present sample and the large resulting uncertainty in $r_0$, a statistical fluke cannot be ruled out.  Alternatively, the unusually luminous quasars in this higher redshift sample may reside in environments whose clustering is strongly ``biased'' with respect to the underlying mass distribution, e.g., at the very highest peaks of the primordial density field.

\section{Conclusion}

Evidence has been presented to substantiate high redshift quasar clustering on scales less than 30 \mpc\ (comoving).  Through $100\,000$ Monte Carlo simulations, the observed quasar separation distribution is found to match a purely random distribution {\em except} for bins smaller than 30 \mpc, where three observed pairs are located.  By scrambling and randomly redistributing the observed quasar coordinates and redshifts, it is shown that the close quasar pairs are not a result of selection effects, and that the scrambled separation distribution is nearly indistinguishable from the Monte Carlo distribution.  Analysis of the minimum quasar pair separations for each of $100\,000$ strip generations reveals that the observed quasar pair separations are significantly smaller than the expected minimum quasar pair separation of 65-75 \mpc.  The nearest neighbor test also tells a tale of high redshift quasar clustering on scales of less than 30 \mpc, showing spikes in the nearest neighbor distance distribution in both the MN and QR strips, while still matching the random Monte Carlo distribution on larger scales.  By assuming a Poisson distribution, the probability of the observed quasar pairs is found to be between 2\% - 5\% each, while the Monte Carlo simulations show that the probability of finding both pairs in the MN strip with separations $\le 30$ \mpc\ is $\sim 0.5$\%.  The correlation function $\xi(r)$ is large and statistically significant for $r<30$ \mpc\ in both strips.

Our results are consistent with those of Kundic (1997), who analyzed the full PTGS quasar sample (we have focused on the high-redshift $z>2.7$ quasars).  Kundic (1997) also finds statistically significant clustering in the quasar sample on scales $r<40$ \mpc, and he finds marginal evidence for an increase in clustering strength towards high redshift.

Assuming $\xi(r) = (r/r_0)^{-1.8}$, we estimate $r_0 = 35 \pm 15$ \mpc\ for these high-redshift quasars, much higher than the correlation length $r_0 \sim 10$ \mpc\ of present-day galaxies, or the correlation length $r_0 \sim 13$ \mpc\ estimated by Shanks \& Boyle (1994) for lower redshift quasars.  While the statistical uncertainties from this sparse sample are considerable, our results suggest that the most luminous objects in the high-redshift universe also exhibit unusually strong spatial clustering.

\acknowledgments

This work was partially supported by National Science Foundation grants AST-9509919 (DPS) and AST-94-15574 (MS).  David Saxe and George Weaver provided valuable hardware and software support for this project.

\end{document}